\documentclass{kluwer}    

\usepackage{epsfig}

\newcommand{\apj}{{\it Astrophys.~J.}}

\newcommand{\Lya}{Ly-$\alpha$\thinspace}
\newcommand{\Msol}{\mbox{M$_\odot$}}
\newcommand{\eg}{e.g.\thinspace}
\newcommand{\ie}{i.e.\thinspace}
\newcommand{\etal}{et al.\thinspace}

\newcommand{\vecr}{{\bf r}}

\newcommand{\vecv}{{\bf v}}
\newcommand{\vecx}{{\bf x}}
\newcommand{\HI}{H{\small I}}
\def\gsim { \lower .75ex \hbox{$\sim$} \llap{\raise .27ex \hbox{$>$}} }
\def\lsim { \lower .75ex \hbox{$\sim$} \llap{\raise .27ex \hbox{$<$}} }
\begin{document}

\begin{article}
\begin{opening}
\title{Numerical Simulations of Galaxy Formation}
\author{Matthias \surname{Steinmetz}\thanks{Alfred P. Sloan Fellow, David and
Lucile Packard Fellow}}
\runningauthor{Matthias Steinmetz}
\runningtitle{Numerical Simulations of Galaxy Formation}
\institute{Steward Observatory\\
University of Arizona\\
Tucson, AZ 85721, USA}
\date{}
\begin{abstract}
The current status of numerical simulations of galaxy formation
is reviewed. After a short description of the main  numerical simulation techniques,
three sample applications illustrate how numerical simulations
have provided deeper insight in the galaxy formation process and how they
have illuminated success and failure of the hierarchical galaxy formation
paradigm: N-body simulations demonstrate that the density profiles of dark
matter halos that form in hierarchical clustering scenarios follow a
characteristic law. A comparison with the kinematics of disk galaxies however
unravels that these density profiles are too concentrated. Hydrodynamical
simulation show that the highly irregular velocity field of merging subclumps
at redshift $z\approx 3$ can easily account for the observed asymmetry in
the absorption profiles of low ionization species in damped \Lya absorption
systems. The built-up of galaxies due to mergers is however also cause for one of
the major inconsistencies of hierarchical structure formation models, the
failure to reproduce the sizes of the present day disk galaxies due to 
excessive transport of angular momentum from the baryonic to the
dark matter component. Hydrodynamical simulations that include star formation
show that scaling laws like the Tully--Fisher relation can readily be reproduced in 
hierarchical scenarios, however the high concentration of dark matter halos
results in a zero-point of the simulated Tully--Fisher relation that is incompatible with observations.
\end{abstract}

\end{opening}

\section{Introduction}

The past couple of years have witnessed a dramatic increase in the quantity and 
quality of observations on the formation and evolution of galaxies. Galaxies are 
routinely identified at redshifts exceeding three and high resolution imaging
allows us to study their internal structure. These data are complemented by high
resolution spectroscopy of QSO absorption systems that provide further clues on 
the evolution of baryons in the universe. In fact this increase has been so
rapid that observations have outgrown their theoretical framework. Traditional
approaches, which rely heavily on the morphological classification of galaxies and 
which intend to disentangle the star formation history of galaxies, seem outdated 
if compared with the much richer structure seen in galaxies at different
redshifts.

Motivated by the increasing body of evidence that most of the mass of the
universe consist of invisible ``dark'' matter, and by the particle physicist's
inference that this dark matter consists of exotic non-baryonic particles, a new 
and on the long run much more fruitful approach has been developed: rather than 
to model the formation and evolution of galaxies from properties of present day
galaxies, it is attempted to prescribe a set of reasonable initial
conditions. The evolution of galaxies is then modeled 
based on physical processes that are considered to be relevant such as gravity,
hydrodynamics, radiative cooling and star formation. The outcome at different
epochs is then confronted against
observational data. 

One scenario that has been extensively tested in that way is the model of
hierarchical clustering, currently the most successful paradigm of structure
formation.  The initial conditions consist of the cosmological parameters
($\Omega, \Lambda, H_0$) and of an initial fluctuation spectrum such as the {\sl cold dark
matter} (CDM) spectrum. The remaining free parameter, the amplitude of these
initial fluctuations, is calibrated by observational data, \eg, the measured
anisotropies of the microwave background.
Structure then grows as systems of progressively larger mass merge and collapse
to form newly virialized systems. This process is illustrated in figure 1,
which shows the hierarchical build-up of a typical galaxy such as the Milky Way.

\begin{figure}
\epsfig{file=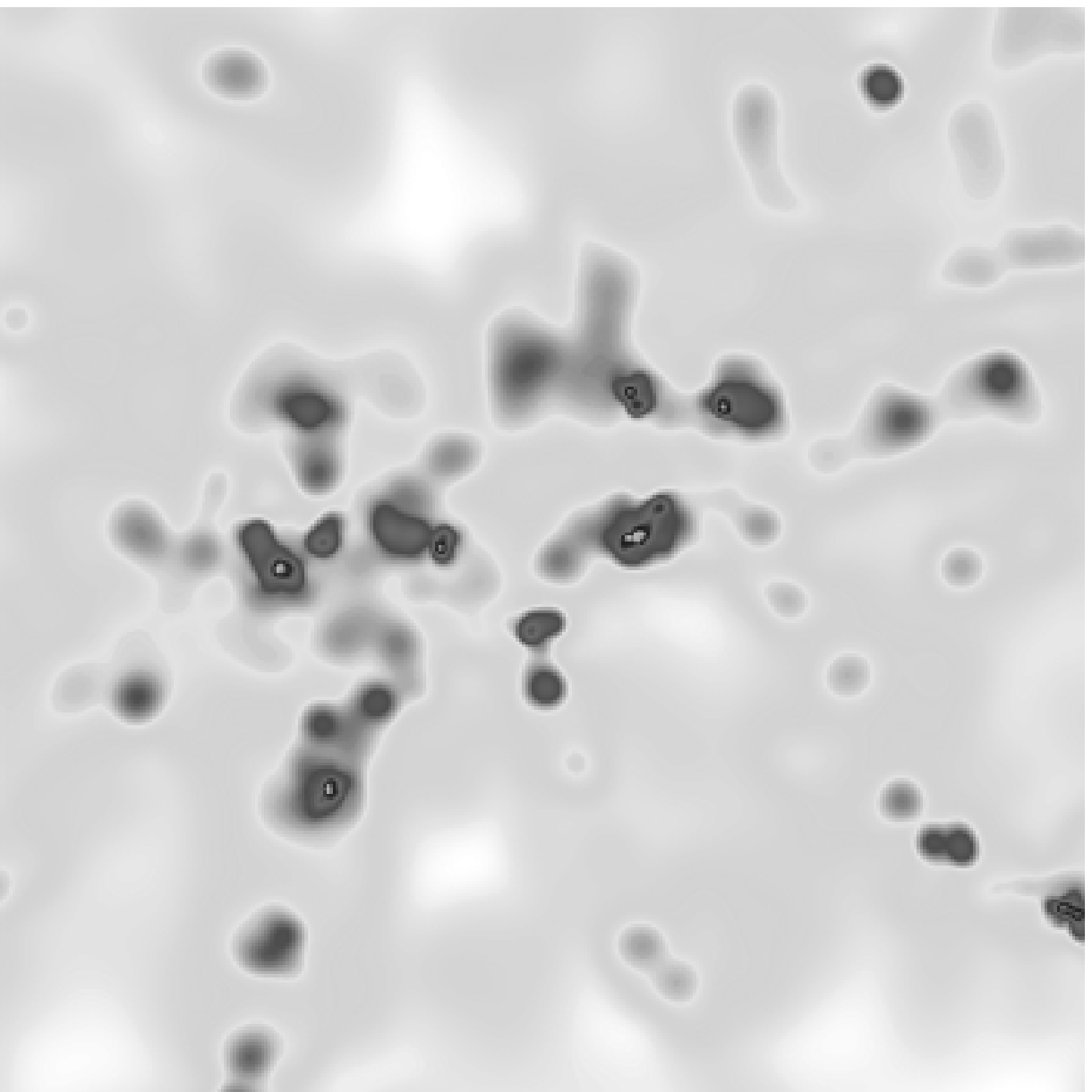,width=3.7cm}\hfill\epsfig{file=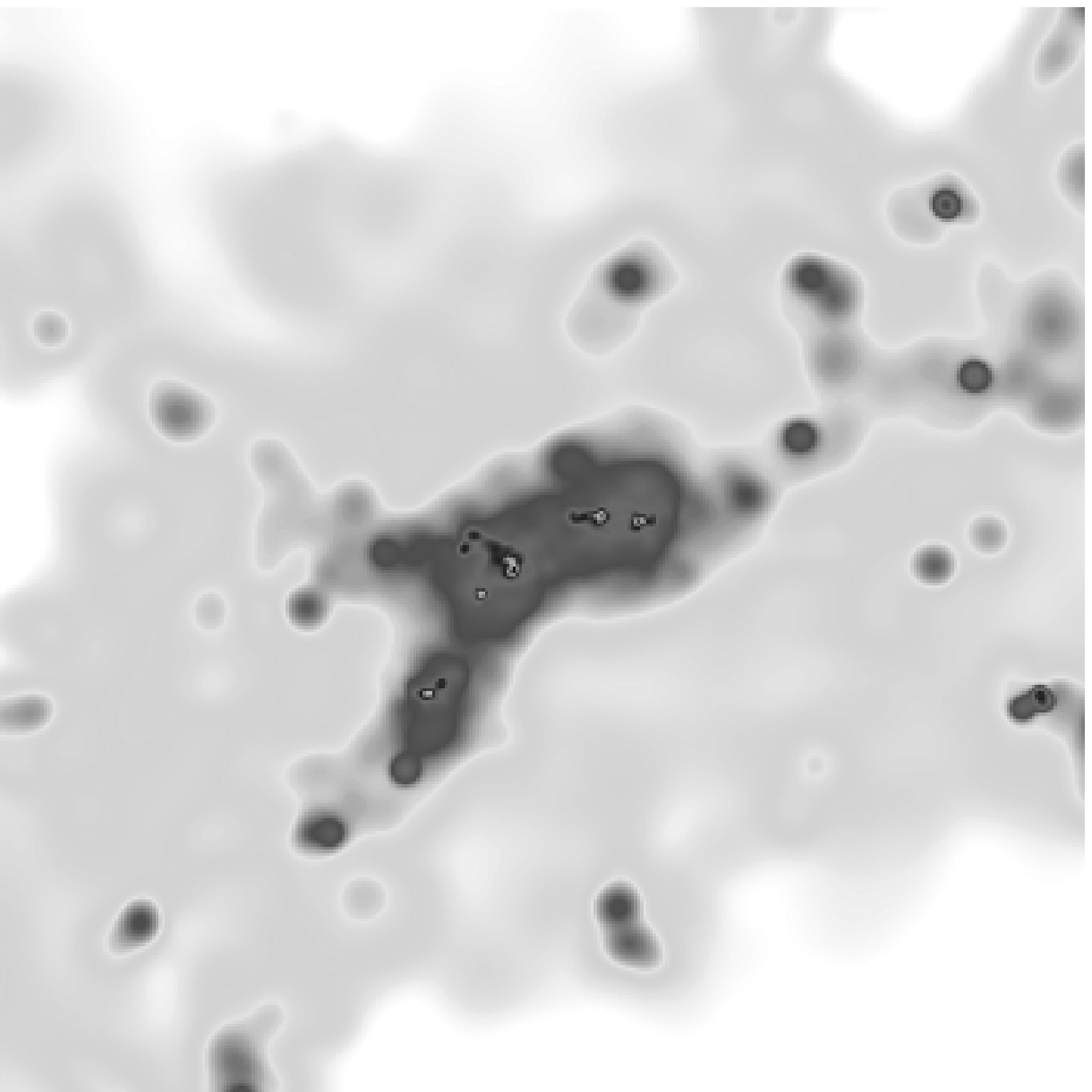,width=3.7cm}\hfill\epsfig{file=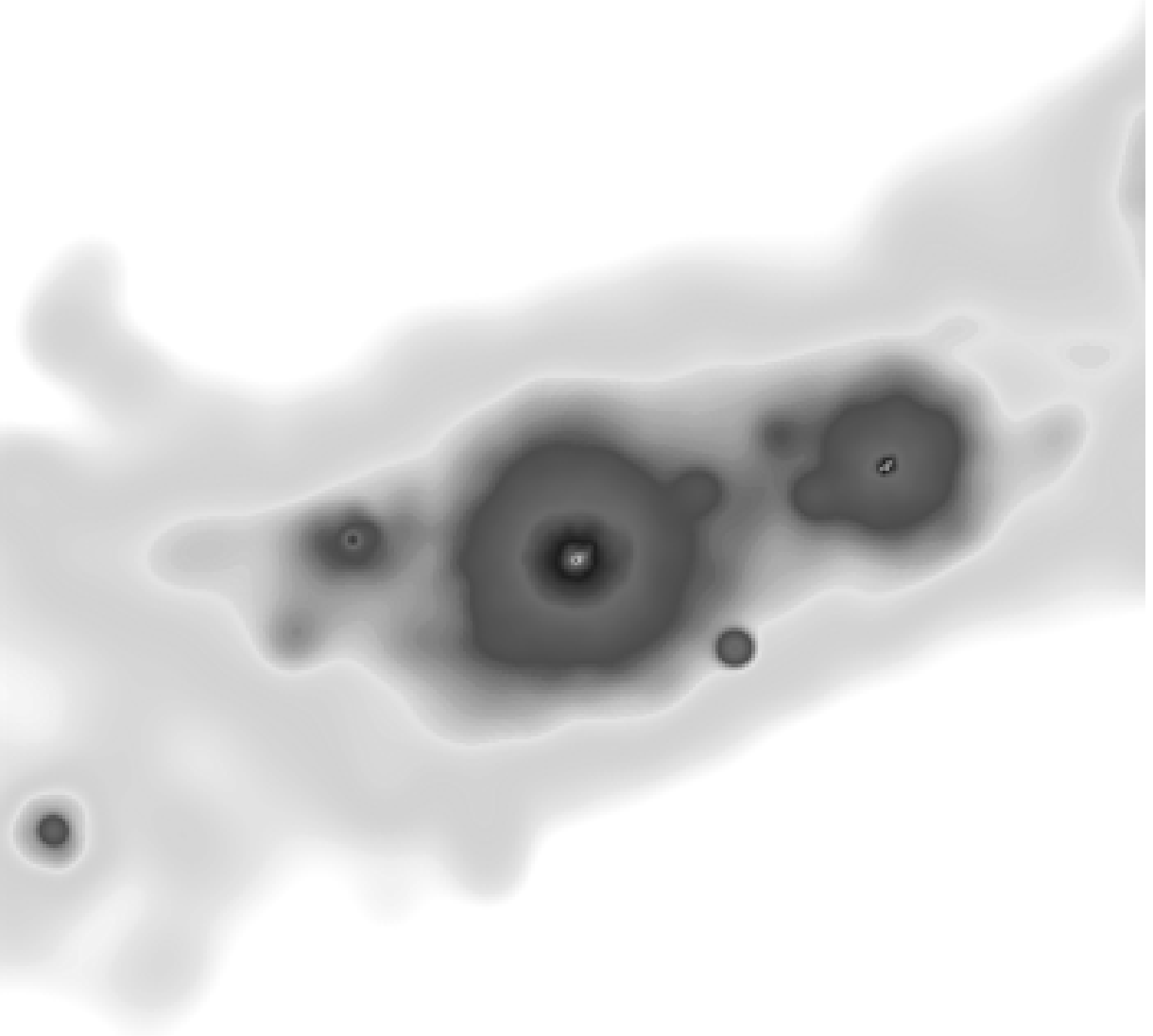,width=3.7cm}
\caption{\label{hier}Hierarchical galaxy formation: Three plots showing the projected
density of a forming galaxy at $z=4$ (left), $z=2$ (middle) and $z=0$ (right). The box
has a side length of 2.8\,Mpc (comoving).}
\end{figure}

These {\sl ab-initio} models can be split in two major categories: (i)
Semianalytical (or phenomenological) models (henceforth SAMs) describe 
the physical processes that affect the formation of 
galaxies by simple, physically motivated recipes
(Kauffmann, Guiderdoni \& White 1994; Cole \etal 1994). Free
parameters are calibrated by comparison with a subset of observational data, \eg 
the galaxy luminosity function. (ii) Numerical simulations model the dynamics
and the interplay of these processes in detail. The boundary between these two
categories is not very well defined. For example, hybrid models are commonly
used in which the large-scale dynamics of the dark matter is modeled by
simulation while SAMs are employed in order to populate dark 
matter halos with galaxies (Kauffmann, Nusser \& Steinmetz 1997). 
Also some commonly used star formation and feedback
prescriptions in hydrodynamical galaxy formation models are not unlike the
recipes used in SAMs (\eg Navarro \& White 1993).

SAMs have proven to be an extremely helpful tool to understand statistical
properties of the galaxy population and how these properties change with
redshift. Due to construction they miss, however, the capability to make strong
predictions on the substructure and detailed morphology of galaxies. It is also
at these galactic length scales (a few kpc) at which the assumptions that enter
semianalytical models and the results of numerical simulations differ the
most. For example, SAMs assume that while galactic halos collapse, gas is heated
up to the virial temperature of the dark matter halo. Simulations however show
that most of the gas is never heated to temperatures exceeding a few
$10^4$\,K. SAMs also assume that the collapse of cooling gas into a rotationally
supported disk proceeds under conservation of angular momentum. Angular
momentum transport between gas and dark matter is neglected, quite in contrast
to the results of numerical simulations. I will come back to this issue in
section 4.

This review intends to report on the state of the art of current
computer simulation and to summarize what major results could be achieved. 
I start this review with an overview of the commonly used simulation
techniques. I will then discuss in detail three examples that illuminate challenge, 
success and failure of numerical simulations of galaxy formation. Such a
selection can only give a glimpse of the large body of results obtained in more
than a thousand refereed publications over the last years addressing topics such
as mass determination of X-ray clusters (Evrard, Metzler \& Navarro 1996),
gravitational lensing by galaxy clusters (Bartelmann \& Steinmetz 1996), 
the physical properties of the \Lya forest (Cen \etal 1994; Hernquist \etal
1996), the formation of first structures in the universe (Abel \etal 1998), 
or the reionization of the universe (Gnedin \& Ostriker 1997), to name only a few.

\section{Techniques}

Over the past three decades quite a variety of numerical simulation techniques
have been developed to simulate the formation of galaxies. These can be roughly
classified into three subgroups: (i) collisionless simulations; (ii)
hydrodynamical simulations; (iii) hydrodynamical simulations that include the
effects of star formation.

\subsection{Collisionless systems}

The distribution function $f(\vecr,\vecv,t)$ of a system consisting of particles 
of mass $m$, which only interact by gravity, can mathematically be described by the Vlasov
equation (collisionless Boltzmann equation):
\begin{equation}
\label{boltz} 
\frac{df(\vecx,\vecv,t)}{dt}  =  \frac{\partial f}{\partial t} + \vecv\cdot\nabla
f -  \nabla\Phi\frac{\partial f}{\partial \vecv} = 0 \\
\end{equation}
The gravitational potential is given by Poisson's equation
\begin{equation}
\label{poisson}
\Delta \Phi = 4\pi\,G\int\,d^3v\,m\,f(\vecx,\vecv,t)\, .
\end{equation}

An astrophysically relevant application of the Vlasov--Poisson system is the
study of systems that only consists of stars and/or (weakly interacting) dark
matter. It turns out, that $N$--body simulations provide a robust and efficient,
although sometimes computationally expensive tool to numerically solve those
equations. 

In
$N$--body--simulations, the trajectories of particles are determined by the laws of
Newtonian dynamics

\begin{eqnarray}
\label{geuler}
\frac{d\vecv_i}{dt} & = & \left. -\nabla \Phi \right|_i\\
\label{grav}
\Phi(\vecr_i) & = & - G \sum \frac{m_j}{|\vecr_j-\vecr_i|}\,\,.
\end{eqnarray}

Every particle of a $N$--body--simulation represents a huge number of
dark matter particles (compare the mass of a body in the simulation
(typically $10^8-10^{12}\,$\Msol) with that of an elementary particle
(100\,GeV)!). $N$--body--simulations can, therefore, be interpreted
as a Monte--Carlo--Approximation of the Vlasov equation, i.e.~the set $(\vecr_i(t),$
$\vecv_i(t), i=1,N)$ samples the distribution function
$f(\vecr,\vecv,t)$. This is a major conceptual difference to N-body
simulations used to model planetary systems or star clusters in which
each particle intends to mimic an actual physical body !

\subsection{Hydrodynamical simulations}

If collisions between particles (\eg ions) can no longer be neglected, 
Vlasov's equation must be replaced by Boltzmann's equation $\frac{d f}{dt} =
[\frac{df}{dt}]_c$, the latter term describing the change in the distribution
function due to collisions. However, if the mean free path of a particle is small compared
to the typical scale of the object under consideration and if the force
between particles is short ranged, a moment expansion of Boltzmann's
equation results in the well know hydrodynamical conservation laws. In so-called
Lagrangean coordinates, \ie coordinates that co-move with a fluid element,
these equations read: 

\begin{eqnarray}
\label{cont}
\frac{d\varrho}{dt} & = & - \varrho \nabla {\bf v}\\
\label{euler}
\frac{d{\bf v}}{dt}  & = & - \frac{1}{\varrho}\nabla P -\nabla\Phi\\
\label{erg}
\frac{d\varepsilon}{dt} & = & - \frac{P}{\varrho}\nabla {\bf v} 
+ \frac{Q - \Lambda(\varrho,T)}{\varrho}\, ,
\end{eqnarray}
where $\varrho$ is the density of the fluid, $\bf v$ its velocity, $P$ 
its pressure and $\varepsilon$ the specific thermal energy, respectively. 
$\Phi$ is the gravitational potential. Equation (\ref{euler}) is usually referred 
to as ``Euler's equation''. The energy equation (\ref{erg}) is derived from the first 
law of thermodynamics. $\Lambda$ and $Q$ denote energy sinks and sources. 
The system is closed by Poisson's equation (\ref{poisson}) and
by an  equation of state, which in the case of an ideal gas reads
\begin{equation}
P= (\gamma-1)\,\varepsilon\,\varrho
\end{equation}
with $\gamma=\frac{5}{3}$. 
\subsection{Hydrodynamical simulations including star formation}
While collisionless and hydrodynamical simulations differ on a very elementary
level, such a distinction cannot be made between simulations that
neglect/include star formation. The only difference at an elementary level is
that in simulations that include the effects of star formation matter, momentum
and energy is exchanged between the collisional (gas) and a collisionless
(stars) component by processes other than gravity. The distinction between
simulation with and without star formation thus appears to be somewhat
arbitrary. I nevertheless consider this discrimination to be valid as a new
element enters: phenomenological modeling. While so far equations were governed
by basic statistical and/or atomic physics, the effects of star formation cannot 
be reduced to such a fundamental level. Actually our understanding of the star
formation process and its interaction with the surrounding interstellar medium
is rather limited.  The best that we can do (at least on the macroscopic scale of a
galaxy) is to empirically parameterize the problem. Most models so far follow
an implementation similar to that outlined by Katz (1992): It is assumed that the 
star formation rate $\frac{d\varrho_*}{dt}$ is proportional to the local gas 
density divided by the local dynamical time scale
\begin{equation}
\frac{d\varrho_*}{dt} = c_*\frac{\varrho_{\rm gas}}{\tau}
\end{equation}
$c_*$ being the star formation efficiency, typically a factor of a few
per cent. Since the local dynamical time scale $\tau$ is proportional to 
$1/\sqrt{G\,\varrho}$, the 
star formation rate grows with the gas density like $\varrho^{1.5}$.
Over a typical star formation time step $\Delta t$, a collisionless 
star particle (which represents of the order of a few million stars) of mass
\begin{equation}
m_* = m_{\rm gas} \left(1 -\exp\left(-\frac{c_*\,\Delta t}{\tau}\right) \right)
\end{equation}
is created and the mass of the gas particle/mesh cell is correspondingly reduced. 
Over the lifetime of a high mass star the supernovae energy is released and the
corresponding mass and energy injected to the gas. In Steinmetz \& M\"uller
(1994, 1995) supernovae also metal enrich the ISM. Simulations exhibit, however, 
that the
energy injected by supernovae affects the dynamics of a galaxy only little. 
Since supernovae energy is
added to gas in the neighborhood of a star forming region, local gas
densities are very high and gas can cool very efficiently (see, however, Yepes
\etal 1997).  
Indeed, Steinmetz \& M\"uller (1995) have shown 
that the net energy
loss, \ie the difference between the energy injected by supernovae and the
energy radiated away due to cooling, is almost identical in simulations with and
without supernova feedback. Navarro \& White (1993) however demonstrated that 
if energy is added to the
kinetic energy, the effect of feedback can be quite drastic. Even in the case of a
low feedback efficiency of about 1\%, the energy released by the first
supernovae is able to strongly suppress further star formation in galaxies with
virial velocities less than about 100\,km/sec.

\subsection{Numerical solution}
The numerical task that has to be performed is to solve a system of partial
differential equations (PDEs), partially of elliptic (\eg Poisson's equation) partially 
of hyperbolic (\eg hydrodynamical equations) character.  In order to solve these PDEs,
quite a number of techniques have been developed that roughly can be split in
two groups:
\begin{itemize}
\item Mesh based methods discretize the PDEs on a mesh and solve the
corresponding difference equation. Most hydrodynamical schemes are mesh based.
\item Particle based methods reformulate (\eg by using kernel estimates) the PDEs
into a set of equations of motion coupled by
interparticle forces. The so-called
{\sl smoothed particle hydrodynamics} (SPH) scheme, the most widely used
hydrodynamical method
in extragalactic astronomy, is such a particle based method. 
\end{itemize}
Both approaches have their disadvantages and their merits and the optimal choice
is problem dependent. As a rule of thumb one can say that mesh based methods are
faster and periodic boundary conditions, which are usually applied for
cosmological simulations, are automatically implemented. 
Mesh based  methods have also been shown to be superior in capturing shocks and 
in dealing with turbulence compared to SPH codes, which rely on artificial viscosity.
The disadvantage is
that the dynamical range is rather limited. Particle methods are computationally
more expensive but have a much larger dynamic range. Furthermore,
periodic boundaries are not automatically provided but have to be added
explicitly, \eg by means of the Ewald summation technique (Ewald 1921;
Hernquist, Bouchet \& Suto 1991). More recently, adaptive mesh refinement (AMR)
techniques (\eg Abel \etal 1998; Kravtsov \etal 1998) have been developed that
combine the merits of particle and mesh based methods, however at the expense 
of a very high code complexity.
Almost all techniques have been adapted to run well on vector- and massively
parallel computers. Particle methods can also run quite efficiently on a
workstation using GRAPE (Sugimoto \etal 1990; Steinmetz 1996), a chip especially designed to
solve the gravitational N-body problem.

\section{N-body simulations -- Density profiles of dark matter halos}

N-body simulation are still the most widely used simulation technique to study
the formation of objects on the largest scales. Most simulations performed in
the seventies and eighties, however, suffered from an insufficient dynamic
range. Those simulations were unable to cover a 
statistically representative volume and to simultaneously resolve individual dark matter
halos. Consequently, cosmological simulation mainly concentrated on the
statistical distribution of matter (\eg determine the two-point correlation
function). In the best case, global properties of dark matter halos such as mass 
and angular momentum could be determined (Barnes \& Efstathiou
1987). Simulations that focussed on the formation of individual galaxies (\eg
van Albada 1982; Barnes \& Hernquist 1992) were performed decoupled from a 
cosmological context.

In recent years, the increasing computer power and the advent of new simulation
techniques allowed to study the formation of an individual object in its full
cosmological context. Multi-mass/AMR simulation techniques 
(Navarro \& Benz 1991; Katz \& White 1993; Bartelmann, Steinmetz \& Weiss 1995) 
resolve individual dark matter halos with several ten thousand particles 
(Navarro, Frenk \& White 1996, 1997), or, more recently, even several million
particles (Moore \etal 1998; Klypin \etal 1998). 

These advanced simulations addressed one of the key problems in
cosmology, the profile of dark matter halos and thus
the shape of the rotation curves of galaxies, one of the major observations that
lead to the postulation of the dark matter dominated universe (\eg Rubin, Thonnard
\& Ford 1978).
Navarro, Frenk \& White (1996, 1997, henceforth NFW)  found that the density profiles of
halos follow a universal law, which can be parameterized by
\begin{equation}
\frac{\varrho(r)}{\varrho_b} = \frac{\delta_n}{\frac{r}{r_s} (1 + \frac{r}{r_s})^2}.
\label{nfw}
\end{equation}
The two parameters are the scale radius $r_s$, which defines the scale where the
profile shape changes from slope $\alpha >-2$ to $\alpha <-2$, and the
characteristic overdensity $\delta_n$. They are related because the mean
overdensity enclosed within the virial radius $r_{\rm vir}$ is $200$.
The slope of the profile $\alpha = d \ln \varrho/d \ln r$ approaches  
$\alpha=-1$ near the halo center  and $\alpha = -3$ at
large radii. At intermediate radii (typically several tens of kpc),
$\alpha\approx -2$ resulting in a flat rotation curve.
Equation~(\ref{nfw}) differs slightly in its asymptotic behavior at large radii
from the profile that was proposed by Hernquist (1990) to describe the mass
profiles of elliptical galaxies.  This profile, which goes like $r^{-4}$ at
large radii, was used by Dubinsky \& Carlberg (1991) to fit the density
distribution of halos that were formed in their CDM-like simulation.

Cole \& Lacey (1996), Tormen, Bouchet \& White (1997), Moore \etal (1998) and 
Huss, Jain \& Steinmetz (1999a), 
among others, have extended the original results of NFW to other initial
spectra, to other cosmologies and to higher spatial and mass resolution. Most of
the above studies show that halos that form in a variety of cosmological models
are well described by equation~(\ref{nfw}). 
These results suggest that the density profile found by NFW is
quite generic for any scenario in which structures form due to hierarchical
clustering, although the slope of the density at small radii is still matter of
some debate (see, e.g., Moore \etal 1998 and Kravtsov \etal 1998).  The power
spectrum and cosmological parameters only enter by determining the typical
formation epoch of a halo of a given mass and thereby the dependence of the
characteristic radius $r_s$ on the total mass of the halo. For {\sl cold dark
matter} type simulations that are normalized to abundance of massive
galaxy clusters, the concentration parameter $c=r_{\rm vir}/{r_s}$ seems to
depend only very little on the actual cosmological parameters $\Omega$ and
$\Lambda$. The initial value given by NFW was $c\approx 10-15$ while recent
higher resolved simulations find this value to be about 40\% higher, 
$c\approx 15-20$ (Moore \etal 1999a; Navarro \& Steinmetz 1999). 

The actual physical mechanisms responsible for such an universal density profile
are as of yet not well understood. Syer \& White (1997) argued that the key to
the NFW profile lies in the hierarchical merging history of cold dark matter
halos. Huss, Jain \& Steinmetz (1999b)
however found that even in hot dark matter simulations and also
in 3D spherically collapse simulations the profiles of dark matter halos are
well fit by a NFW profile.  Similar
results have also been obtained by Moore \etal (1999a) for a warm dark matter
model. Huss \etal (1999b) argue that the universal profile is a more generic 
feature of gravitational
collapse in a cosmological setting and is related to the angular momentum
distribution of halos. The mechanism by which angular momentum is acquired
can vary, potential mechanism include merging (for hierarchical models) and
radial orbit instabilities (for spherical collapse models).

\begin{figure}
\mbox{\epsfig{file=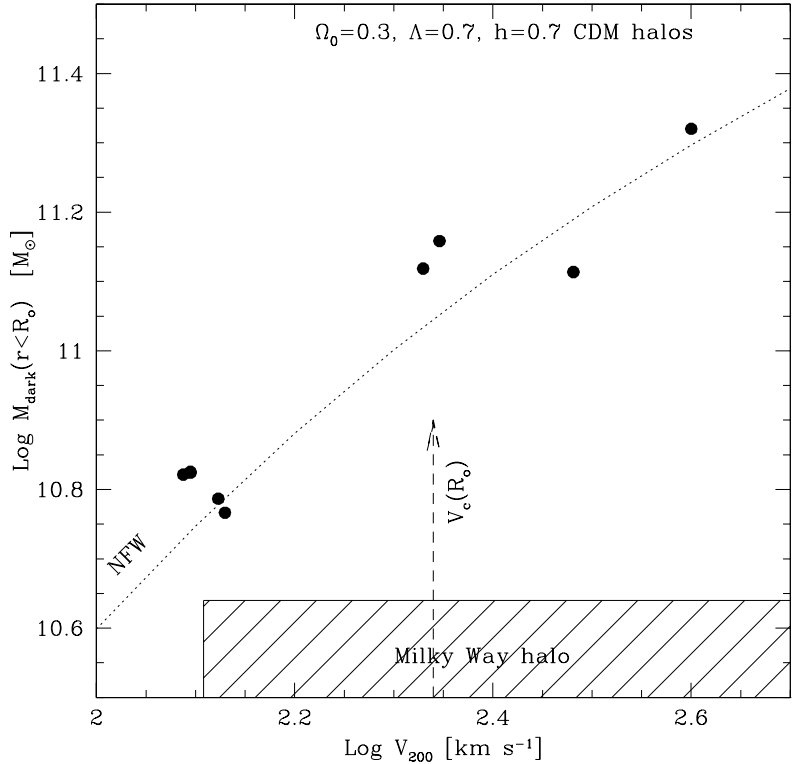,width=11cm}}
\caption{\label{milky}Dark mass enclosed within a radius $R_{o}= 8.5$ kpc, the
Sun's distance from the center of the Milky Way, versus the circular velocities
of $\Lambda$CDM halos. The shaded region highlights the allowed parameters of
the dark halo surrounding the Milky Way, as derived from observations of
Galactic dynamics and by assuming that the disk mass cannot exceed the total
baryonic content of the halo. The filled circles show the loci of $\Lambda$CDM
halos as determined from high-resolution N-body simulations. The solid line is
the circular velocity dependence of the dark mass expected inside $R_{o}$
for halos that follow the density profile proposed by NFW. The
circular velocity dependence of the NFW ``concentration'' parameter of the
simulated halos is well approximated on these scales by $c \approx 20 \,
(v_{\rm vir}/100 {\rm \, km \, s}^{-1})^{-1/3}$ (dotted line). }
\end{figure}

The existence of such an universal density profile makes strong predictions on
the shape of the rotation curve of galaxies that can be confronted against
observations. Such a test on galactic scales is particularly interesting since
few observational constraints on such small scales have been used to adjust the
parameters of the different flavors of the cold dark matter scenario. 
Such a comparison reveals that the asymptotic slope of the density profile 
at small radius of $\alpha = -1$,
seems at odds with rotation curve studies
of dark matter-dominated galaxies (Moore 1994; Flores \& Primack 1994; McGaugh
\& De Block 1998; Moore \etal 1999a; Navarro 1999). 
Unfortunately, the scales where deviations are most
pronounced (the inner few kpc) are also the most compromised by numerical
uncertainties (most simulations relevant to this problem published to date have
gravitational softening scales of order 1 kpc). The comparison is thus
rather uncertain. For example, Kravtsov \etal (1998) have argued, on the basis
of simulations similar to those used by the other authors, that CDM halos are
actually {\it consistent} with the rotation curves of dark matter-dominated
disks, a somewhat surprising result that illustrates well, nonetheless, the
vulnerability of numerical techniques on scales close to the numerical
resolution of the simulations. 

In order to circumvent these problems, Navarro \& Steinmetz (1999, henceforth NS99) used, 
rather than the dark matter density profile near
the center, the {\it total amount} of dark mass within the main body of
individual galaxies. For spiral galaxies, this criterion implies that
simulations that can estimate reliably the amount of dark mass within a couple
of exponential scalelengths may be safely used for comparison with
observations. For bright spirals like the Milky Way this corresponds to radii of
about $5$-$10$ kpc, well outside the region that may be compromised by numerical
artifacts in the current generation of N-body experiments.

The argument to constrain the total amount of dark matter within the solar
circle $R_0$ goes as follows:

\begin{itemize}
\item The total stellar mass of the Milky Way puts a lower limit on the total
baryonic mass enclosed within the virial radius of the Milky Way's dark matter
halo. Since the baryon to dark matter ratio within the virial radius is the same 
as the cosmologically representative value (White \etal 1993), this puts a lower 
limit $v_{\rm vir, min}$ on the mass and thus on the circular velocity of the Milky Way's
halo. Using N-body simulations, a lower limit $M_{\rm DM,lim}$ on the mass within the solar
circle can be deduced.
\item The rotation velocity of the Milky Way can be translated into a total mass 
enclosed within the solar circle. By subtracting the observed stellar mass an
upper limit on the observationally inferred mass within the solar circle
$M_{\rm DM,est}$ can be given.
\item In order to be consistent with the observations, $M_{\rm DM,lim}$ must be
smaller than the $M_{\rm DM,est}$ at least for some halos with 
$v_{\rm vir} > v_{\rm vir,min}$. 
\end{itemize}
The main difficulty in this argument is a proper estimate of potential
observational uncertainties. This analysis has been done in NS99, the results
are shown in figure 2, which compares, as a function of halo mass, the
dark mass estimate $M_{\rm DM,est}$ with the results of simulations of several
$\Lambda$CDM halos.  Halo masses ($M_{\rm vir}$) are measured inside the
radius, $r_{\rm vir}$, of a sphere of mean density 200 times the critical
density for closure, and are typically characterized by the circular velocity at
that radius.  The comparison shows clearly a major discrepancy between the
maximum dark matter inside $R_{o}$ allowed by observations and the results of
the numerical experiments. For example, $\Lambda$CDM halos with circular
velocities similar to that of the Milky Way disk ($v_{\rm vir} \approx
v_{\rm rot}(R_{o})=220$ km s$^{-1}$) have about {\it three times} more dark mass inside
the solar circle than inferred from observations. Even for the extreme case
where the halo has the strict minimum circular velocity  $v_{\rm vir,min}$, the
simulations indicate an excess of more than $50\%$ in the dark mass within
$R_{o}$.

This serious discrepancy only worsens if some extra
dark material is drawn inside $R_{o}$ by the formation of the disk. A rough
estimate of the magnitude of this correction can be made by assuming that the
halo responds adiabatically to the assembly of the disk; the discrepancy then
increases from $50\%$ to almost $80\%$ for the least massive halo. Halos formed 
in the $\Lambda$CDM scenario are too
centrally concentrated to be consistent with observations of the dynamics of the
Galaxy.

The results thus confirm the problem already seen in the rotation curves of disk
galaxies, that dark matter halos as they form in a cold dark matter scenario are
too concentrated. A related problem has been recently published by Moore \etal
(1999a,b) and by Klypin \etal (1999) who demonstrate that cold dark matter
scenarios predict that the
local group should host 
fifty
times as many dwarf galaxies than actually
observed. In order to reconcile the cold dark matter scenario, a power spectrum
would be required that suppresses power on galactic and subgalactic scales while
keeping the large scale properties of the model virtually unchanged.  This would
in principle allow galaxy-sized dark halos to collapse later and thus become
less centrally concentrated. One problem that afflicts such a modification is
that they may hinder the formation of massive galaxies at high redshift, at odds
with the mounting evidence that such galaxies are fairly common at $z \,
\gsim \, 3$ (see, e.g., Steidel \etal 1998).

\section{Hydrodynamical simulations -- The kinematics of damped \Lya systems}

The major advantage of hydrodynamical simulations compared to N-body simulations is
that they model the dynamics of the {\sl visible} matter. Although
hydrodynamical simulations were considerably successful
in understanding the details of the galaxy formation
process, the largest impact so far is related to the properties of QSO
absorption systems where numerical simulations can
explain the basic properties of QSO absorbers
covering many orders of magnitude in column density (Cen \etal 1994; Hernquist
\etal 1996; Zhang, Anninos \& Norman 1995). 
Indeed, hydrodynamical simulation were responsible for a
paradigm shift as QSO absorbers are no longer considered to be caused by
individual gas clouds. Absorbers of different column density (\Lya forest,
metal line systems, Lyman limit systems and damped \Lya absorption systems) are rather
reflecting different aspects of the large scale structure of the universe. 
While the lowest column density systems ($\log N \approx 12-14$) 
arises from gas in voids and sheets of the ``cosmic web'', systems of higher
column density are produced by filaments ($\log N \approx 14-17$) or even by gas
that has cooled and collapsed in virialized halos ($\log N > 17$).  

\begin{figure}
\mbox{\epsfig{file=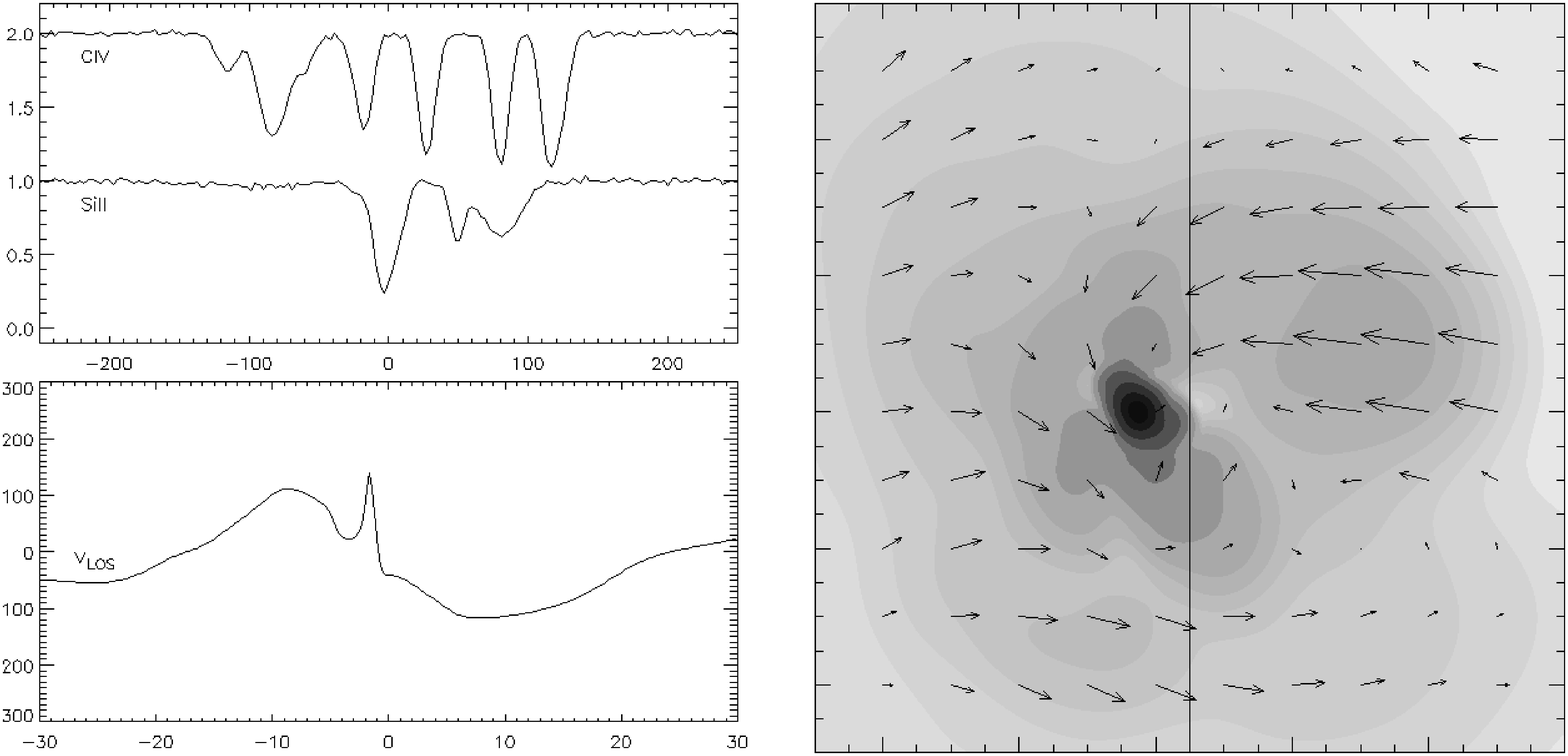,width=12cm}}
\caption{\label{damped}Right: Color map of the column density distribution in a
60\,kpc around a damped system. Black correspond to \HI\ densities $\log n (\HI)
> 1.5$), light grey to $\log n (\HI) \approx -3$). White arrows indicate the
velocity field. The white line correspond to the line-of-sight(LOS). In the
lower left plot, the velocity
field along the LOS is shown. The upper left plot shows the absorption line in
CIV 1548 (top) and SiII 1808 (bottom). For readability, CIV has been displaced
by 0.5 in flux.}
\end{figure}

The kinematics of damped \Lya absorption systems (DLAS) at high redshift serves
as a very nice example to demonstrate how oversimplifying assumption may lead to
wrong implications on the validity of a cosmological model 
and to show how the full numerical treatment can avoid those
artificial contradictions.  DLASs have often been interpreted as large,
high-redshift progenitors of present-day spirals that have evolved little apart
from forming stars (Wolfe 1988). Kauffmann (1996), however, studied the evolution 
of DLASs in the CDM structure formation scenario in which disks form by continuous
cooling and accretion of gas within a merging hierarchy of dark matter halos and
found that the total cross section
was dominated by disks with comparably low rotation velocities (typically 70
km/sec).  Prochaska \& Wolfe (1998) however, observed much larger velocity
spreads (up to 200\,km/s) and came to the conclusion that only models in which
the lines-of-sight (LOS) intersects rapidly rotating large galactic disks can
explain both the large velocity spreads and the characteristic asymmetries of
the observed low ionization species (\eg, SiII) absorption profiles, in strong
contradiction to the prediction of the (semianalytical) cold dark matter model.

So how may a numerical simulation solve this problem ? The critical assumption
that enters the semianalytical models is that DLASs are equilibrium
disks. Numerical simulations (Haehnelt, Steinmetz \& Rauch 1998) show, however,
that this is a poor assumption and that asymmetries and non-equilibrium effects
play an important role. Figure 3 shows a typical configuration that
gives rise to a high redshift DLAS with an asymmetric SiII absorption profile.
The velocity width of about 120\,km/s is also quite similar to typical
observations.  However, no large disk has yet been developed and also the
circular velocity of the collapsed object is only 70\,km/s.  The physical
structures that underly DLASs are turbulent gas flows and inhomogeneous
density structures related to the merging of two or more clumps, rather than
large rotating disks similar to the Milky Way. Rotational motions of the gas
play only a minor role for these absorption profiles.  A more detailed analysis
also demonstrates that the numerical models easily pass the statistical tests
proposed by Prochaska and Wolfe, \ie, hierarchical clustering, in particular the
CDM model, is consistent with the kinematics of high-$z$ DLASs.

\begin{figure}
\mbox{\epsfig{file=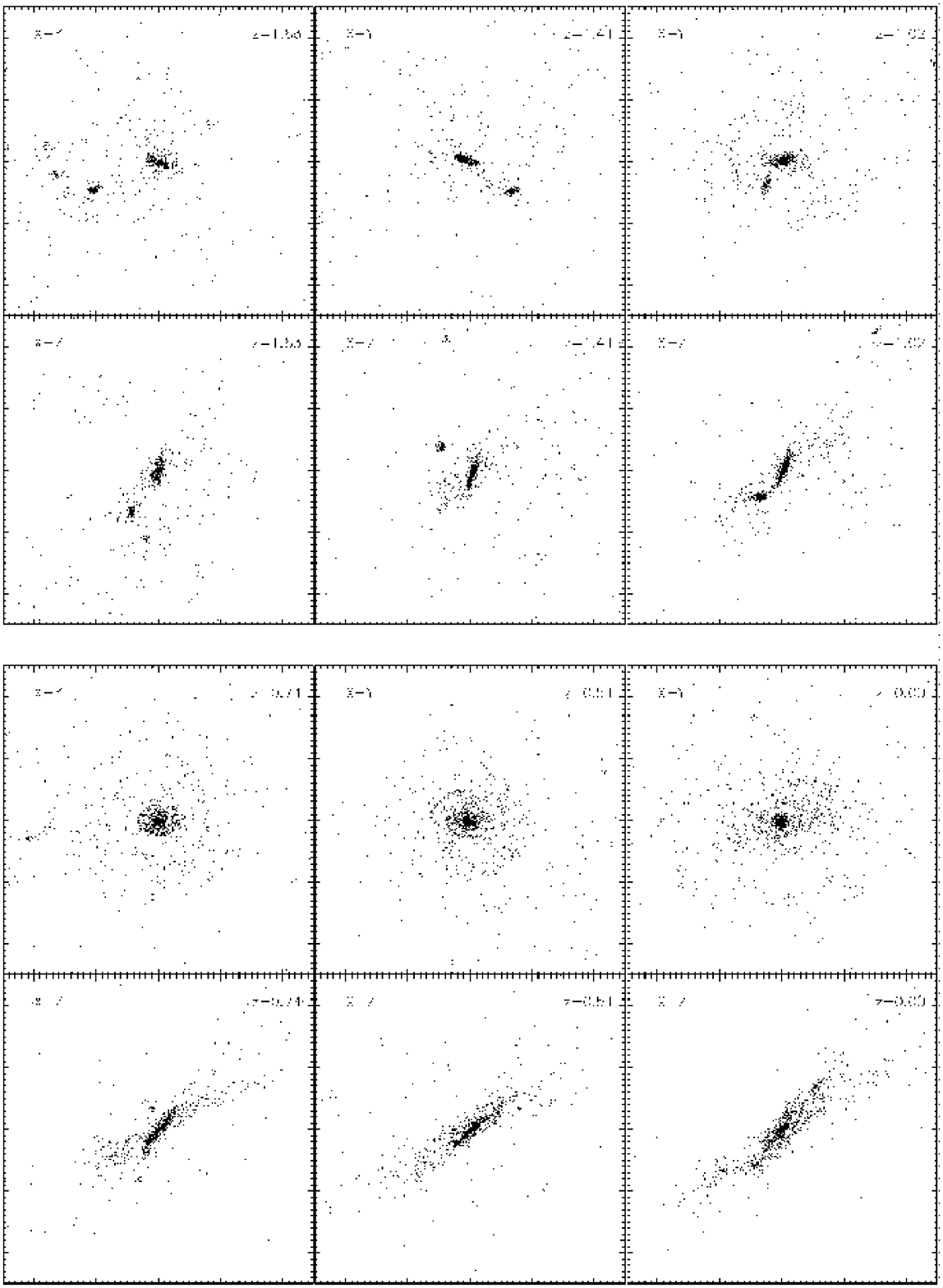,width=12cm}}
\caption{\label{disk}The distribution of gas projected in the X-Y 
and Y-Z plane shown for 6 different redshifts.
The gas infall is mainly lumpy. Diffusely infalling gas settles 
to form a rotationally supported disk.}
\end{figure}

While the irregular structure of a galaxy in the process of formation was
helpful to alleviate the problem concerning the kinematics of DLASs, one may
wonder whether it hurts the assembly of large galactic disks at low redshifts as 
major mergers are usually associated with the transformation of spirals into
ellipticals (for a review, see Barnes \& Hernquist 1992). Figure 4
indicates that on a qualitative level this seems not to be a problem. Although
the gas is accreted in a fairly lumpy manner, at redshift $z=0$ a nice disk like 
structure has developed.

\begin{figure}
\mbox{\epsfig{file=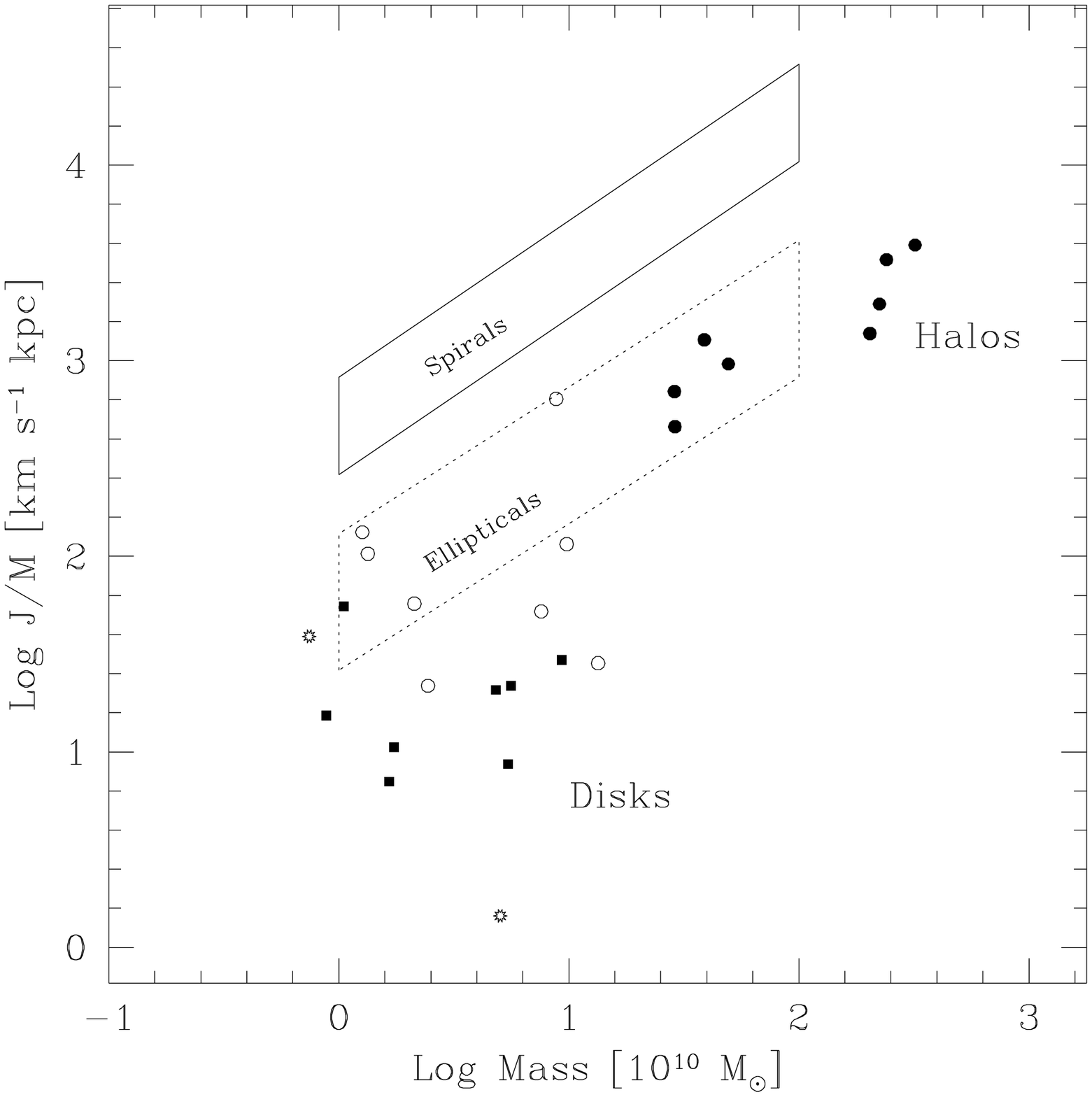,width=6cm}\epsfig{file=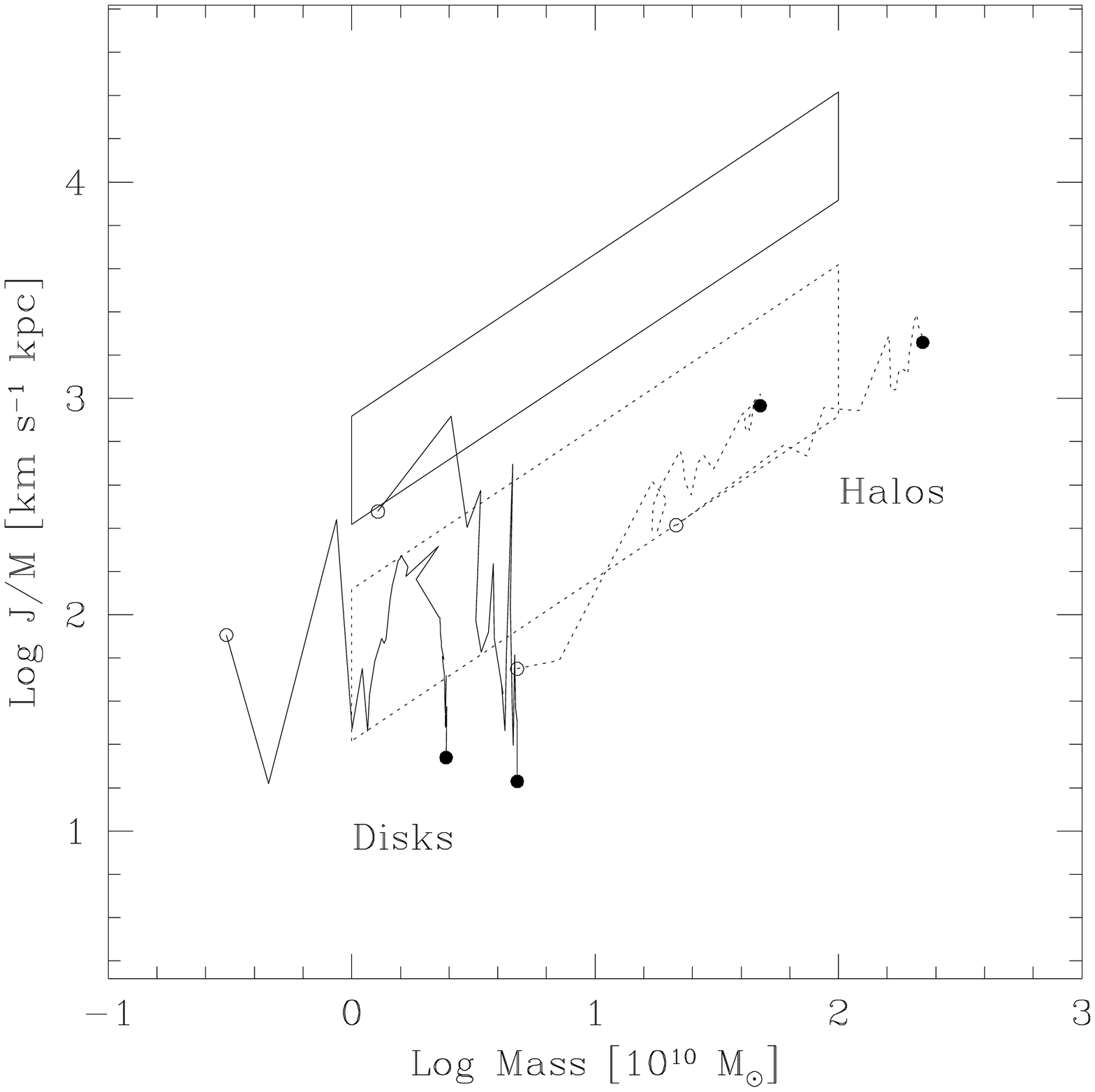,width=6cm}}
\caption{\label{angmom}The specific angular momentum of dark halos and gaseous
disks, as a function of mass. The boxes enclose the region occupied by spiral
and elliptical galaxies, as given by Fall (1983). Open circles, solid squares
and starred symbols correspond to the specific angular momenta of gaseous disks,
solid circles for the hosting dark matter halos.  Right: Evolution of the dark
halo and central gaseous disk in the $J/M$ versus $M$ plane, from $z=5$ (open
circles) to $z=0$ (solid circles).  The mass of the system grows steadily by
mergers, which are accompanied by an increase in the spin of the halo and a
decrease in the spin of the central disk. The latter results from angular momentum being
transferred from the gas to the halo during mergers.}
\end{figure}

However, a closer look at the detailed structure reveals a major shortcoming of
these disks, they are too concentrated (Navarro \& Benz 1991; Navarro, Frenk
\& White 1995; Navarro \& Steinmetz 1997). This, however, is not a problem in
semianalytical models, which relatively easily can reproduce the sizes of present
day galaxies. One may again ask: What is the difference between the
semianalytical model (Mo, Mao \& White 1997) and the numerical simulations, in
particular since they are based on the same structure formation model ? And
again the reason can be found in the assumptions that enter the semianalytical
models: Semianalytical models assume that gas collapses under conservation of
angular momentum (Fall \& Efstathiou 1980), an assumption that, as will be
shown, is only very poorly fulfilled.

Figure 5 (left) shows the specific angular momentum of dark
matter halos and of their central gaseous disks at $z=0$, as a function of
mass.  If, as suggested by Fall \& Efstathiou
(1980), the collapse of gas would proceed under conservation of angular
momentum, the baryonic component would have the same specific angular momentum $J/M$
as the dark matter, however, its corresponding mass would be a factor of 20
smaller (for $\Omega_{\rm bary} = 0.05$, $\Omega_0=1$). These disks would be
located only slightly below the box for spiral galaxies. However, figure 5
demonstrates clearly that the spins of gaseous disks are about an
order of magnitude lower than that. This is a direct consequence of the
formation process of the disks. Most of the disk mass is
assembled through mergers between systems whose own gas component had previously
collapsed to form centrally concentrated disks. During these mergers, and
because of the spatial segregation between gas and dark matter, the gas
component transfers most of their orbital angular momentum to the surrounding
halos. While the specific
angular momentum of dark matter halos increases with decreasing redshift, that of gaseous
disk decreases (figure 5 (right)). 

\section{Hydrodynamical simulations including star formation -- The Tully
Fisher relation}

\begin{figure}
\epsfig{file=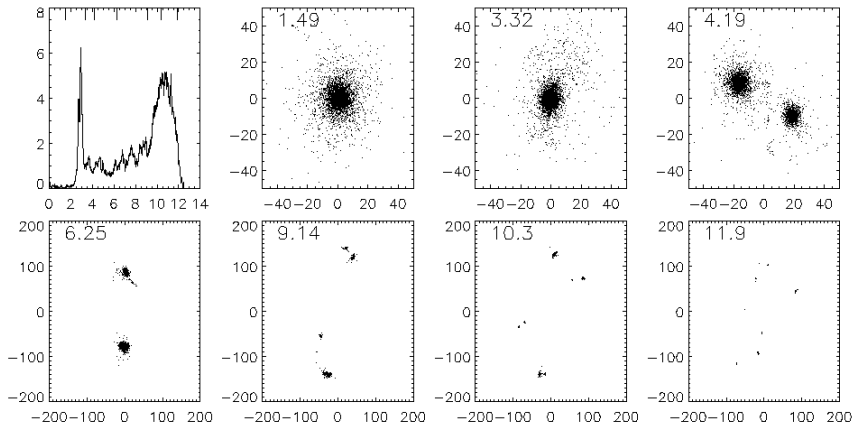,width=11.5cm}
\caption{\label{sfhist}Upper left: Star formation rate (in \Msol/yr) versus lookback time of a 
forming galaxies (circular velocity $\approx 100\,$km/sec. 
Other plots: seven snapshots of the star particle
distribution of the forming galaxy. Lookback time for each snapshot is given in 
the upper left corner.}
\end{figure}

Some of the problems mentioned in the former sections may be alleviated by
including star formation and related feedback processes such as stellar winds and
supernovae (see \eg Weil, Eke \& Efstathiou, 1998). 
The major stumbling block designing such a simulations is the
choice of a reasonable star formation recipe and the calibration of free
parameters. The simulations presented in this section were performed using the star
formation recipe as described in section 2.3. The star formation efficiency was
calibrated such that a Kennicutt type relation between gas surface density and
star formation can be reproduced (Kennicutt 1998). 

Figure 6 illustrates the formation history of such a galaxy: The star 
formation rate peaks at early times (lookback time $\approx 11\,$Gyr) due to the nearly
simultaneous collapse of a number of halos that later on merge to form two
spiral galaxies. The star formation rate in each progenitor rarely exceeds $\sim 1
M_{\odot}$/yr, although its combined rate can reach about 8 $M_{\odot}$/yr. The
star formation rate in the two galaxies, which have formed by $t_{\rm lookback} =
7\,$Gyr, is fairly constant and slowly declining. The two galaxies merge at
$t_{\rm lookback} = 3\,$Gyr. Due to the merger gas streams to the center in a
fashion similar to that described in Mihos \& Hernquist (1994) where it is
rapidly transformed into stars. The resulting ``star burst'' consumes almost all 
the gas and star formation at $t_{\rm lookback} < 2\,$Gyr is essentially
quenched, resulting in the formation of an elliptical galaxy. 
Galaxies that do not experience such a major merger at later epoch
(not shown) continue to slowly transform gas into stars.

The success of such a model can be further assessed by testing, to what extent such a
model can reproduce scaling relations that
link total luminosity, rotation speed, and angular momentum of
disk galaxies such as the Tully--Fisher (TF) relation.  Figure 7 shows the results of such an
investigation,  
the simulated $I$-band TF relation at $z=0$ for two cosmological
scenarios, a standard CDM ($\Omega=1, \Lambda=0$) and a $\Lambda$CDM
($\Omega=0.3, \Lambda=0.7$) scenario. The simulated TF relation is compared
with the data of Giovanelli et
al.~(1997), Mathewson, Ford \& Buchhorn (1992) and Han \& Mould (1992).  The
slope and scatter of the simulated TF relation are in fairly good agreement with
the observational data.  This results also holds in other bandpasses: the model
TF relation becomes shallower (and the scatter increases) towards the blue, just
as in observational samples (see Steinmetz \& Navarro 1999). 
The model TF relations are also very tight. In the I-band the rms scatter is only
0.25 mag, even smaller than the observed scatter of $\sim 0.4$ mag. This must be
so if the results are to agree with observations: scatter in the models reflects
the intrinsic dispersion in the TF relation, whereas the observed scatter
includes contributions from both observational errors and intrinsic
dispersion. If, as it is usually argued, both effects contribute about equally
to the observed dispersion in the TF relation, then the intrinsic scatter in the
I-band should be comparable to the $\sim 0.25$ mag found in the models.

The zero-point is, however, in serious disagreement with the data, the simulated 
TF relation being  almost two magnitudes too faint at given rotation speed. 
One may argue that the discrepancy is related to the particular star formation
and feedback parameterization, however very similar results can be obtained
for quite different star formation prescriptions (Navarro \& Steinmetz, in
preparation). 

\begin{figure}
\mbox{\epsfig{file=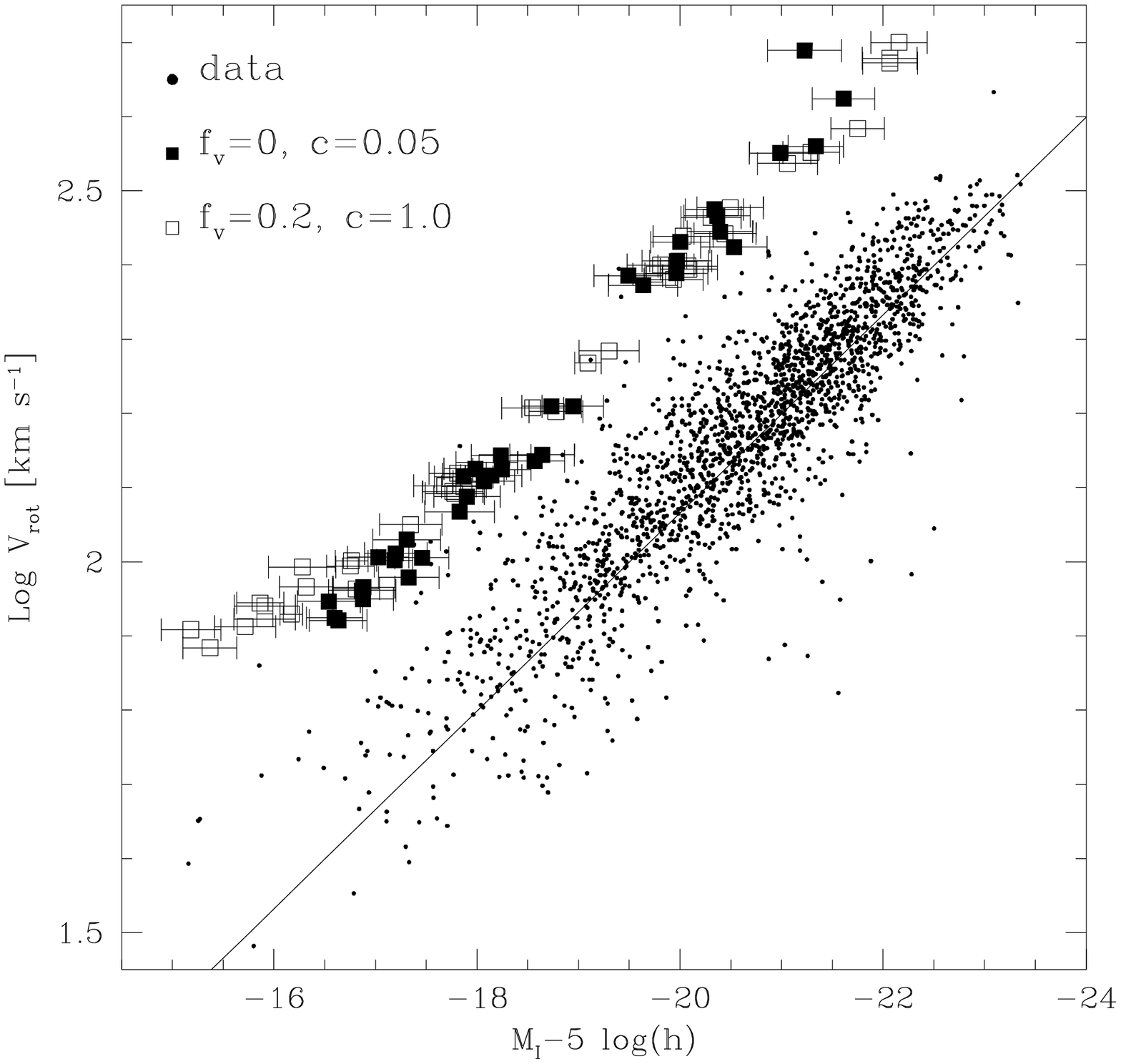,width=5.5cm}\hskip1cm\epsfig{file=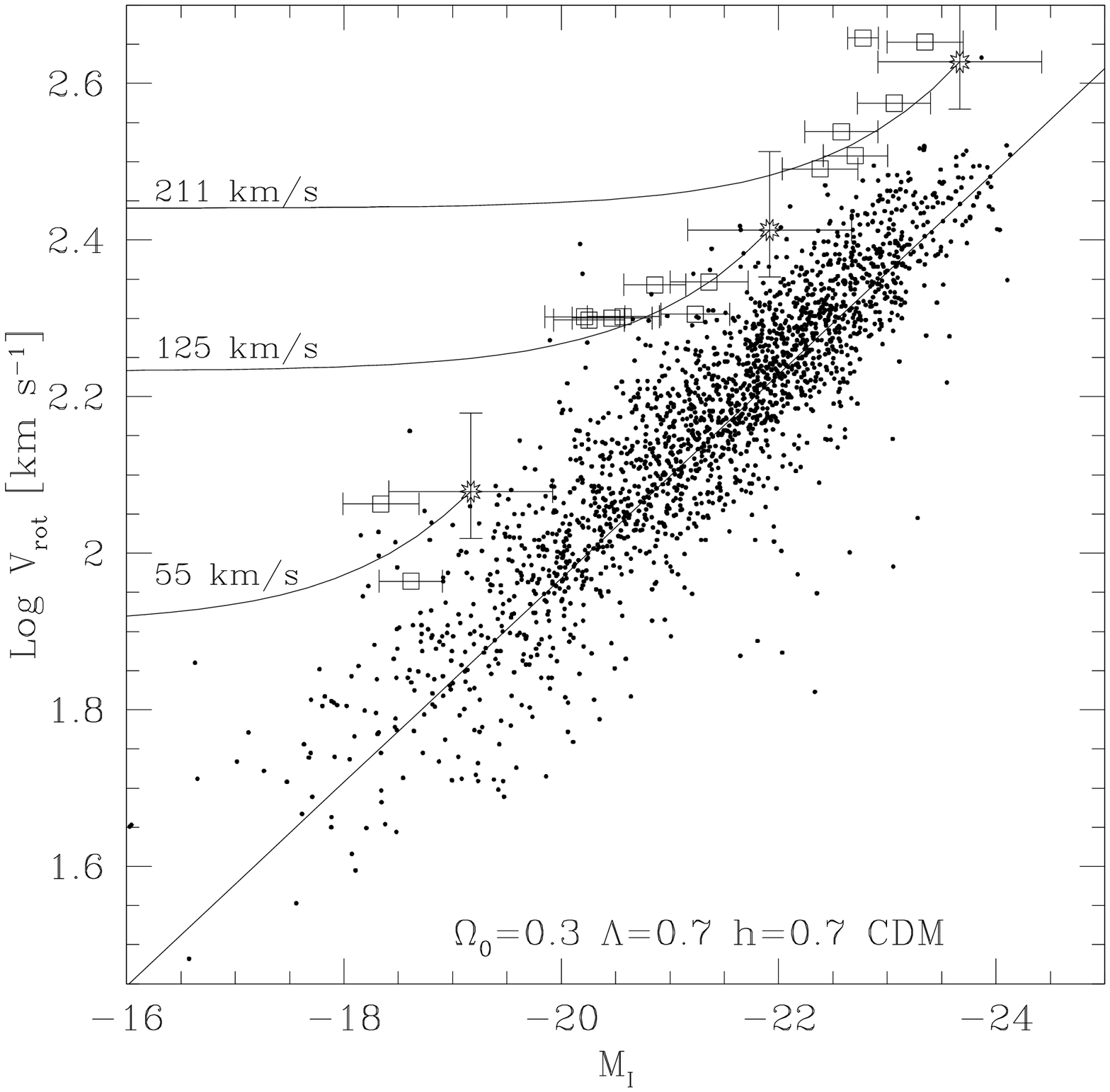,width=5.5cm}}
\caption[\label{tully}]{Left: I-band TF relation at $z=0$ for an
$\Omega=1,\Lambda=0$ CDM scenario. The error bars in
the simulated data span the difference in magnitudes that results from adopting
a Salpeter or a Scalo IMF. Right: I-band TF relation at $z=0$ for 
$\Omega=0.3,\Lambda=0.7$. See text for details.}
\end{figure}

In order to further understand the physical origin of the success and failure of 
the TF modelling, I will focus on  the following three questions: (i) Why is the 
slope in such good agreement ? (ii) Why is the scatter so small, in particular
considering that the variation in the star to dark matter ratio changes quite
substantially between individual halos ? (iii) What causes the offset in the
zero--point of the TF relation ?

The first question is rather straight forward to answer (see \eg Mo, Mao \&
White 1998): The slope is just a moderate modulation of the $M\propto v^3$
scaling of dark matter halos that reflects the approximately scalefree process
of assembly of collisionless dark matter into collapsed, virialized systems.

So why is the scatter so small although the star to gas ratio between individual
halos varies by a factor of three to five ? The answer is related to the
dependence of the rotation velocity of the galactic
disk to the star to DM ratio. As more and more baryon assemble in the central stellar disk, their
luminosity increases but also the disk rotation velocity due to the gravity of
the additional matter near the center. This effect is amplified as the additional gravity also
pulls dark matter towards the center. The solid-line curves in figure
7 (right) illustrate this effect for 
three representative dark halos of different mass (or luminosity, since 
a constant mass-to-light ratio of $\Upsilon_I=2$ is assumed).  In each case
the total disk mass varies from zero to $M_{\rm disk,max}$, the maximum value
compatible with the baryonic content of the halo, the rightmost point of each
curve. As the disk mass increases,
each hypothetical galaxy moves from left to right across the plot. When the disk
mass becomes comparable to the dark mass inside $2.2 \, r_{\rm disk}$ the curve
inches upwards and becomes essentially parallel to the observed TF
relation. Consequently a variation in the star to dark matter fraction results
in a shift parallel to the TF relation and thus does not cause a
substantial additional scatter.

It is also clear from figure 7 why the models fail to reproduce the
observed zero-point of the TF relation. Even under the extreme assumption that galaxies
contain {\it all} available baryons in each halo, simulated disks are almost two
magnitudes fainter than observed. Increasing the baryonic mass of a halo has
virtually no effect on this conclusion, since in this case model galaxies would
just move further along paths approximately parallel to the TF
relation, as shown in figure 7. Disk galaxies assembled inside CDM
halos therefore cannot match the observed TF relation.

Perhaps the most uncertain step in this argument is the stellar mass-to-light
ratio adopted for the analysis. The horizontal ``error bar'' shown on the
starred symbols in figure 7 indicates the effect of
varying the $I$-band mass-to-light ratio by a factor of two from the fiducial
value of $2$ in solar units. This is not enough to restore agreement with
observations, which would require $(M_{\rm disk}/L_I) \sim 0.4$, a value much
too low to be consistent with standard population synthesis models. The vertical
``error bars'' illustrate the effect of varying the ``concentration'' of each
halo by a factor of two. Even with this large variation in halo structure, the
model disks fail to reproduce the observations.

The mismatch of the zero-point of the TF relation is closely linked to the
failure to reproduce the dynamics of the Milky Way discussed earlier. Because
the dark halos are quite centrally concentrated, the assembly of a massive galaxy at
the center raises $v_{\rm rot}$ above and beyond the halo circular velocity, by
up to $60 \%$. The only way to collect a massive disk galaxy without increasing
$v_{\rm rot}$ significantly over $v_{\rm vir}$ is to have dark halos that are less
centrally concentrated, consistent with the conclusion of section 3.

\section{Summary and Conclusions}

I presented results on recent efforts to model the formation and evolution of
galaxies in a hierarchical structure universe using high resolution computer
simulations. I demonstrated that only numerical simulation can take full account
of the dynamic of the formation process and the complicated interplay between
different physical processes such as, \eg, accretion and merging, star formation
and feedback, photo heating and radiative cooling. Observational data can easily 
be misinterpreted if these effects are not properly included.  For example, the
apparent inconsistency of hierarchical structure models and the kinematics of
high-$z$ damped \Lya absorption systems could be easily solved by properly
accounting for the complicated non-equilibrium dynamics of galaxies in the
process of formation. Further successes include:

\begin{itemize}
\item Gaseous and stellar disks resembling spiral galaxies can readily be produced
in numerical simulations. Merging of spiral galaxies can trigger star burst and
result in the formation of ellipticals. 
\item Adopting star formation recipes that match observational constraints (\eg
Kennicutt's law), the slope and scatter of the TF can be easily
reproduced in such numerical simulations.
\end{itemize}

However, numerical simulations also unraveled a number of serious
inconsistencies of currently favored structure formation models:

\begin{itemize}
\item Structure formation models such as the different flavors of the cold
dark matter scenario make strong predictions on the profiles of dark matter
halos. However, models such as the $\Lambda$CDM model predict
dark matter halos that are too concentrated to be consistent with the
rotation curves of disk galaxies, the kinematics of the Milky Way and the zero
point of the TF relation. 

\item The probably most serious problem of hierarchical clustering
scenarios is related to the angular momentum of disk galaxies. Tidal torques,
which spin up dark matter halos early on in their formation history, only provide 
marginally enough angular momentum to explain the sizes of 
present day disk galaxies. Maintaining the hierarchical build-up of galaxies and 
simultaneously avoiding substantial exchange  of angular momentum from the
gas to the dark matter due to mergers appears to be a major challenge.
\end{itemize}

But the probably even more important result is that numerical simulations have
demonstrated to be capable of making strong, falsifiable predictions on the
structure of galaxies at length scales of several kpc and smaller. 
The new 8m-class telescopes and proposed space missions will be able to probe
this exact length scale in high-z galaxies. Numcerical simulations thus provide
an indispensable tool for establishing the theoretical framework within which
these new observations can be interpreted.

\begin{acknowledgements} 

This article includes work from collaborations with
M.~Haehnelt, A. Huss, B.~Jain, J.~Navarro and M.~Rauch. 
This work has been supported by the National
Aeronautics and Space Administration under NASA grant NAG 5-7151, by the
National Science Foundation under NSF grant 9807151, and by fellowships from
the Alfred P.~Sloan Foundation and the
David and Lucile Packard foundation. Travel support to participate the
conference was provided by the Anglo American Chairman's Fund, and SASOL.
I also acknowledge the
hospitality of the Max-Planck Institut f\"ur Astronomie and of the 
Max-Planck Institut f\"ur Astrophysik, where parts of this manuscript
have been written.
\end{acknowledgements}

\end{article}

\begin{thebibliography}{}
%
\bibitem{a1}Abel, T., Anninos, P., Norman, M.L., Zhang, Y., 1998, ApJ, 508, 518.
%
\bibitem{a2}van Albada, T. S., 1982, MNRAS, 201, 939.
%
%
\bibitem{b1}Barnes, J., Efstathiou, G., 1987, ApJ, 319, 575.
%
\bibitem{b2}Barnes, J., Hernquist, L., 1992, ARA\&A, 30, 705.
%
\bibitem{b3}Bartelmann, M., Steinmetz, M., 1996, MNRAS, 284, 431.
%
\bibitem{b4}Bartelmann, M., Steinmetz, M., Weiss, A., 1995, A\&A, 297, 1.
%
\bibitem{c1}Cen, R., Miralda-Escud\'e, J.,  Ostriker, J.P., Rauch, M., 1994, ApJ, 437, 9.
%
\bibitem{c2}Cole, S.M., Arag\'on-Salamanca, A., Frenk, C.S., Navarro, J.F., Zepf,
S.E., 1994, MNRAS, 271, 781.
%
\bibitem{d3}Cole, S., Lacey, C., 1996, MNRAS, 281, 716.
%
\bibitem{d0}Dubinski, J., Carlberg, R.G., 1991, ApJ, 378, 496
%
\bibitem{e1}Evrard, A.E., Metzler, C.E., Navarro, J.F., 1996, ApJ, 469, 494.
%
\bibitem{e2}Ewald, P.P., 1921, Ann.~Phys., 64, 253.
%
\bibitem{f1}Fall, S.M., 1983, in {\it Internal Kinematics and Dynamics of Galaxies}, Athanassoula E. (ed.), (Dordrecht: Reidel), p. 391.
%
\bibitem{f2}Fall, S.M., Efstathiou, G., 1980, MNRAS, 193, 189.
%
\bibitem{F3}Flores, R.A., Primack, J.R., 1994, ApJ, 427, L1.
%
317, 595.
%
\bibitem{g1} Giovanelli, R., Haynes, M.P., Herter, T., Vogt, N.P., 1997,
AJ, 113, 22.
%
\bibitem{g2}Gnedin, N.Y., Ostriker, J.P., 1997, ApJ, 486, 581.
%
\bibitem{h1}Han, M., Mould, J.R., 1992, ApJ, 396, 453.
%
\bibitem{h2}Haehnelt, M., Steinmetz, M., Rauch, M., 1998, ApJ, 495, 647.
%
\bibitem{h3}Hernquist L., 1990, ApJ, 356, 359.
%
\bibitem{h4}Hernquist, L., Bouchet, F., Suto, Y., 1991,ApJS, 75, 231.

\bibitem{H5}Hernquist, L., Katz, N., Weinberg, D.H., Miralda--Escude J., 1996, ApJ, 457, L51.
%
\bibitem{h6}Huss A., Jain B., Steinmetz M., 1999a, MNRAS, in press.
%
\bibitem{h7}Huss A., Jain B., Steinmetz M., 1999b, ApJ, 517, 64.
%
\bibitem{k1}Kauffmann, G., 1996, MNRAS, 281, 475.
%
\bibitem{k0}Kauffmann, G., Nusser, A., Steinmetz, 1997, MNRAS, 286, 795.
%
\bibitem{k2}Kauffmann, G., White, S.D.M., Guiderdoni, B., 1994, MNRAS, 267, 981.
%
\bibitem{k3}Katz, N., 1992, ApJ, 391, 502.
%
\bibitem{k4}Katz, N., White, S.D.M., 1993, ApJ, 412, 455.
%
\bibitem{k10}Kennicutt, R.C., 1998, ARA\&A, 36, 189.
%
\bibitem{k5}Klypin, A., Gottl\"ober, S., Kravtsov, A.V., Khokhlov, A.M., 1999, ApJ, 516, 530.
%
\bibitem{k6}Klypin, A., Kravtsov, A.V., Valenzuela, O., Prada, F., 1999, ApJ,
552, 82.

%
\bibitem{K7} Kravtsov, A.V., Klypin, A.A., Bullock, J.S., Primack, J.R.,
1998, ApJ, 502, 48.
%
\bibitem{m1}Mathewson, D.S., Ford, V.L., Buchhorn, M., 1992, ApJS, 81, 413
%
\bibitem{M2}McGaugh, S.S., De Block, W.J.G. 1998, ApJ, 499, 41.
%
\bibitem{m3}Mihos, C., Hernquist, L., 1994, ApJ, 431, 9.
%
\bibitem{M4}Mo, H.J., Mao, S., White, S.D.M., 1998, MNRAS, 295, 319.
%
\bibitem{M5}Moore, B., 1994, Nature, 370, 629.
%
\bibitem{m6}Moore, B., Governato, F., Quinn, T., Stadel, J., Lake, G., 
1998, ApJ, 499, 5.
%
\bibitem{m7}Moore, B., Quinn, T., Governato, F., Stadel, J., Lake, G., 
1999a, MNRAS submitted, (astro-ph/9903164).
%
\bibitem{m8}Moore, B., Ghigna, S., Governato, F., Lake, G., Quinn, T., Stadel, J., 
1999b, ApJ (Letters), in press (astro-ph/9907411).
%
\bibitem{n1}Navarro, J.F., 1999, ApJ, submitted (astro-ph 9807084).
%
\bibitem{n2}Navarro, J.F., Benz W., 1991, ApJ, 380, 320.
%
\bibitem{n3}Navarro, J.F., Frenk, C.S., White, S.D.M., 1995, MNRAS, 275, 56.
%
\bibitem{n4}Navarro, J.F., Frenk, C.S., White, S.D.M., 1996, ApJ, 462, 563.
%
\bibitem{n5}Navarro, J.F., Frenk, C.S., White, S.D.M., 1997, ApJ, 490, 493 (NFW). 
%
\bibitem{N6}Navarro, J.F., Steinmetz, M., 1997, ApJ, 478, 13.
%
\bibitem{N7}Navarro, J.F., Steinmetz, M., 1999, ApJ, in press (astro-ph/9908114) (NS99).
%
\bibitem{n8}Navarro, J.F., White, S.D.M., 1993, MNRAS, 265, 271.
%
\bibitem{p1}Prochaska, J. X., Wolfe, A. M. 1998, ApJ, 507, 113. 
%
\bibitem{r1}Rubin, V.C., Thonnard, N.,  Ford, V.K., 1978, ApJ, 225, L107.
%
\bibitem{s1} Steidel, C.C., Adelberger, K.L., Dickinson, M., Giavalisco,
M., Pettini, M., Kellogg, M., 1998, \apj, 492, 428.
%
\bibitem{S2}Steinmetz, M., 1996, MNRAS, 276, 1005.
%
\bibitem{s3}Steinmetz, M., M\"uller, E., 1994, A\&A, 281, L97.
%
\bibitem{s4}Steinmetz, M., M\"uller, E., 1995, MNRAS,  276, 549.
%
\bibitem{S5}Steinmetz, M., Navarro, J.F., 1999, ApJ, 513, 555.
%
\bibitem{s6}Sugimoto, D., Chikada, Y., Makino, J., Ito, T.,
Ebisuzaki, T., Umemura, M., 1990 {\it Nature} {\bf 345} 33
%
\bibitem{s7}Syer, D., White, S.D.M., 1998, MNRAS, 293, 337.
%
\bibitem{t1}Tormen, G., Bouchet, F.R., White, S.D.M., 1997, MNRAS, 
286, 865.
%
\bibitem{w1} Weil, M.L., Eke, V.R., Efstathou, G., 1998, MNRAS, 300,773. 
%
\bibitem{w2} White, S.D.M., Navarro, J.F., Evrard, A.E., \& Frenk, C.S., 1993, Nature,
366, 429.
%
\bibitem{w3} Wolfe, A.M., 1988, in {\it QSO Absorption Lines: Probing the Universe},
Proc. of the QSO Absorption Line Meeting, Baltimore, 1987, Cambridge 
University Press.
%
\bibitem{y1}Yepes, G., Kates, R., Khokhlov, A., Klypin, A., 1997, MNRAS, 284, 235.
%
\bibitem{z1}Zhang, Y., Anninos, P., Norman, M.L., 1995, ApJ, 453, L57.
%
\end{thebibliography}
\end{document}